\documentclass[letterpaper]{article} % DO NOT CHANGE THIS
\usepackage{aaai2026}  % DO NOT CHANGE THIS
\usepackage{times}  % DO NOT CHANGE THIS
\usepackage{helvet}  % DO NOT CHANGE THIS
\usepackage{courier}  % DO NOT CHANGE THIS
\usepackage[hyphens]{url}  % DO NOT CHANGE THIS
\usepackage{graphicx} % DO NOT CHANGE THIS
\usepackage{natbib}  % DO NOT CHANGE THIS AND DO NOT ADD ANY OPTIONS TO IT
\usepackage{caption} % DO NOT CHANGE THIS AND DO NOT ADD ANY OPTIONS TO IT
\usepackage{amsmath}
\usepackage{amssymb}
\usepackage{multirow}
\usepackage{array}
\usepackage{booktabs}
\usepackage{algorithm}
\usepackage{algorithmic}

\usepackage{newfloat}
\usepackage{listings}
\DeclareCaptionStyle{ruled}{labelfont=normalfont,labelsep=colon,strut=off} % DO NOT CHANGE THIS
\floatstyle{ruled}
\newfloat{listing}{tb}{lst}{}
\floatname{listing}{Listing}
\nocopyright % -- Your paper will not be published if you use this command
\title{HPR3D: Hierarchical Proxy Representation for High-Fidelity 3D Reconstruction and Controllable Editing}
\author{
    Tielong Wang,
    Yuxuan Xiong,
    Jinfan Liu,
    Zhifan Zhang,
    Ye Chen,
    Yue Shi,
    Bingbing Ni
}
\affiliations{
    Shanghai Jiao Tong University\\
    Shanghai, China\\
}

\begin{document}

\maketitle

\begin{abstract}
Current 3D representations like meshes, voxels, point clouds, and NeRF-based neural implicit fields exhibit significant limitations: they are often task-specific, lacking universal applicability across reconstruction, generation, editing, and driving. While meshes offer high precision, their dense vertex data complicates editing; NeRFs deliver excellent rendering but suffer from structural ambiguity, hindering animation and manipulation; all representations inherently struggle with the trade-off between data complexity and fidelity. To overcome these issues, we introduce a novel 3D Hierarchical Proxy Node representation. Its core innovation lies in representing an object's shape and texture via a sparse set of hierarchically organized (tree-structured) proxy nodes distributed on its surface and interior. Each node stores local shape and texture information (implicitly encoded by a small MLP) within its neighborhood. Querying any 3D coordinate's properties involves efficient neural interpolation and lightweight decoding from relevant nearby and parent nodes. This framework yields a highly compact representation where nodes align with local semantics, enabling direct drag-and-edit manipulation, and offers scalable quality-complexity control. Extensive experiments across 3D reconstruction and editing demonstrate our method's expressive efficiency, high-fidelity rendering quality, and superior editability.
\end{abstract}

\section{Introduction}
% 3D 表征非常重要
% 传统的 3D 表征难以实现多粒度的可控编辑。主要原因是它们缺乏多层级的架构。
% 为了解决这个问题，我们提出了一个表征，它包含多层级的控制点，xxx。
% 通过操纵控制点，我们能够对目标物体实现多层级的可控编辑。
% 我们的贡献点包括：1. 一个全新的由多层级控制点组织成的表征；2. 一个自适应构建多层级控制点的高效算法；3. 实验表面，我们的结果在重建和编辑等任务上能够取得 sota 的结果。

% In recent years, rapid advancements in 3D vision have catalyzed substantial progress in a wide range of tasks, including 3D generation, reconstruction, and editing. Central to these developments is the choice of 3D representation, which serves as a foundational element of the 3D processing pipeline and is often tailored to the requirements of specific tasks. Broadly, existing 3D representations can be classified into two categories: rendering-oriented representations and surface reconstruction-oriented representations.
Recent progress in 3D vision has advanced tasks like generation, reconstruction, and editing. A key factor underlying these developments is the choice of 3D representation, which forms the foundation of the processing pipeline and is typically tailored to specific task requirements. Broadly, existing representations fall into two categories: rendering-oriented and surface reconstruction-oriented.

Rendering-oriented representations take multi-view images as input and aim to synthesize accurate renderings from novel viewpoints. Prominent examples include neural implicit representations~\cite{mildenhall2021nerf, barron2021mipnerf, mueller2022instant, kaizhang2020, liu2020neural, yu2021plenoctrees, reiser2021kilonerf, martin2021nerf, pumarola2021d, Oechsle2021ICCV, chen2022tensorf} and 3D Gaussian Splatting (3DGS)~\cite{kerbl20233d, Huang2DGS2024}, both of which leverage differentiable rendering to learn from 2D data, bypassing traditional stereo reconstruction.

While implicit representations excel at view synthesis, they are not directly compatible with 3D mesh formats widely used in VR, film, and gaming~\cite{Shi_2021_ICCV, hui2022neural}. These representations distort Euclidean space and lack correspondence to semantic parts, making editing and structural understanding difficult. Existing editing methods~\cite{yuan2022nerf, park2021nerfies, athar2022rignerf, xiong2025dualnerftextdriven3dscene, wang2023seal3d} often involve complex constraints without reliable precision.

3DGS has recently gained traction for its ability to model fine details via spatially varying Gaussian densities~\cite{kerbl20233d, hamdi2024ges, lin2024vastgaussian, yu2024mip, liang2024gs, yan2024multi, jiang2024gaussianshader, li2024mipmap}. However, it remains a point-based representation lacking surface continuity. Although recent work attempts to extract smooth surfaces from 3DGS~\cite{Huang2DGS2024, guedon2023sugar, yu2024gaussian, wolf2024gsmesh}, challenges remain in quality, editability, and structural control.

Surface reconstruction-oriented representations form another key class of 3D representations, aiming to directly model surface geometry for high-precision reconstruction. Signed Distance Function (SDF)-based methods~\cite{wang2021neus, neus2, Park_2019_CVPR, icml2020_2086, NeuralPull, li2021d2im, Yu2022MonoSDF, yariv2021volume} learn a continuous signed distance field using neural networks, with the zero-level set implicitly defining the object’s surface. High-quality meshes can then be extracted via isosurface extraction.

To improve flexibility and accuracy, recent approaches store SDF values in dense volumetric structures such as tetrahedral meshes~\cite{shen2021deep} or voxel grids~\cite{shen2023flexible}, decoupling representation from network inference. While often achieving better geometric fidelity, these methods suffer from significant storage and computation costs. The resulting mesh representations support precise, local editing by manipulating vertices or facets in selected regions. However, such operations are typically labor-intensive and require point-wise control. Even with techniques like cage-based deformation~\cite{jakab2021keypointdeformer} or As-Rigid-As-Possible (ARAP) editing~\cite{ARAP_modeling:2007}, it remains difficult to isolate local edits without inadvertently affecting surrounding areas, limiting their applicability to relatively simple deformations.

To address these limitations, we introduce a novel hierarchical proxy representation.
Given a 3D object in an arbitrary modality (\emph{e.g.}, mesh, point cloud, implicit field, or multi-view images), our approach constructs a hierarchical tree of control points to represent the underlying geometry in a compact and structurally coherent manner. To facilitate high-fidelity texture reconstruction, we associate multi-scale texture features with control points at each level and employ a learned texture decoder conditioned on this hierarchical representation.
By explicitly encoding both geometry and appearance across multiple levels of abstraction, our method enables precise, part-aware local editing at varying levels of granularity—functionality that is challenging to realize with existing representations.

% 最近，有一篇同期工作 Mash~\cite{li2025mash} 也提出了一种利用控制点+局部特征来表征物体表面的方法。他们用存储在 anchor points 上的球谐函数来表征目标物体表面的局部 patch，实现了高表征精度和高存储压缩率。然后，MASH 的每个表面 patch 都仅被一个 anchor point 控制，这意味着 MASH 重建出的表面本质上是很多小 patch 的“简单拼接”。单独编辑任何一个 anchor point 都会导致 MASH 表征的表面出现裂缝。这意味着 MASH 表征即使使用了控制点来表征目标物体，他们依旧无法通过编辑控制点的方法来实现快速、简便的编辑操作。
Recently, a concurrent work, MASH~\cite{li2025mash}, also proposes a control-point-based surface representation that leverages localized features. Specifically, it encodes local surface patches using spherical harmonics stored at anchor points, achieving high representation accuracy and compression efficiency.
However, in MASH, each surface patch is exclusively controlled by a single anchor point, resulting in a surface that is essentially a simple concatenation of many independently represented patches. As a consequence, editing any individual anchor point typically leads to visible discontinuities or seams on the reconstructed surface.
This limitation indicates that, although MASH incorporates control points in its representation, it lacks support for intuitive and flexible editing via control point manipulation. In contrast, our method enables smooth, consistent, and multi-scale geometry editing through hierarchical proxy points.

Our contributions are threefold:
\begin{itemize}
    \item We propose a novel hierarchical proxy representation that enables compact and accurate modeling of target objects, while supporting controllable geometry and texture editing at multiple levels of granularity.
    \item We introduce an efficient and adaptive algorithm for constructing the hierarchical control point structure, allowing for fast and semantically coherent organization.
    \item Extensive experiments demonstrate that our approach achieves state-of-the-art performance on tasks such as reconstruction and editing.
\end{itemize}

\section{Related Works}

\subsection{3D Representations}

Existing 3D representations can be broadly classified into two categories: rendering-oriented and surface-oriented representations, depending on their primary design objectives and downstream applications.

Rendering-oriented representations aim to synthesize photorealistic images from novel viewpoints and typically adopt neural implicit representations. Neural Radiance Fields (NeRF)~\cite{martin2021nerf, barron2021mipnerf, mueller2022instant, kaizhang2020, SHI_DARF, yu2021plenoctrees, reiser2021kilonerf, pumarola2021d, Oechsle2021ICCV, chen2022tensorf} represent volumetric radiance and density fields via neural networks and render views through differentiable volume rendering. Although NeRF-based methods achieve impressive view synthesis quality, they often produce low-fidelity geometry due to the lack of explicit surface constraints, and their implicit nature makes localized editing difficult~\cite{liu2021editing, yuan2022nerf, park2021nerfies, athar2022rignerf, shi2023separated3d, wang2023seal3d}.

Recently, 3D Gaussian Splatting (3DGS)~\cite{kerbl20233d} has emerged as a fast and high-quality alternative, modeling scenes with spatially distributed Gaussians. While effective for rendering, 3DGS lacks topological structure, limiting its use for geometry reconstruction and editing. Several follow-up methods~\cite{guedon2023sugar, Huang2DGS2024, yu2024gaussian, wolf2024gsmesh} aim to extract surfaces from 3DGS, but the resulting meshes remain coarse and difficult to manipulate due to the underlying point-based nature.

In contrast, surface-oriented representations focus on modeling geometry directly. Signed Distance Function (SDF)-based methods~\cite{wang2021neus, neus2, Park_2019_CVPR, icml2020_2086, NeuralPull, li2021d2im, Yu2022MonoSDF, yariv2021volume} define surfaces as the zero level-set of a learned field and extract meshes using Marching Cubes~\cite{lorensen1998marching}. These approaches achieve high geometric accuracy but remain difficult to edit due to the entangled nature of the implicit field.

To improve editability and reconstruction fidelity, recent methods~\cite{shen2021deep, shen2023flexible} store SDF values on dense tetrahedral or voxel grids and use differentiable mesh rendering. While more flexible than purely neural fields, such dense volumetric structures incur high memory costs and offer limited support for fine-grained or hierarchical editing.

\subsection{Mesh Editing Methods}
% 传统的 mesh 几何编辑通常是直接对其 vertices, edges, 及 faces 进行直接的编辑。这种编辑方法是十分消耗人工与时间的。相比于直接操纵 mesh 中的元素，free-form deformation 编辑方法，例如基于 lattice~\cite{10.1145/800031.808573} 或 cage~\cite{jakab2021keypointdeformer}，将空间划分为多个区间，使得用户能够通过拖拽 keypoints，实现对 mesh 的局部进行更快速和简洁的编辑。然而这类方法只能进行单一尺度的空间划分，无法同时实现多尺度下的精准局部编辑。另一类 handle-based editing，例如 Laplacian Editing~\cite{10.1145/1057432.1057456} 和 As-Rigid-As-Possible (ARAP)~\cite{ARAP_modeling:2007} deformation，通过指定几个控制点，在局部区域发生形变时尽量保持局部特性不变。这类方法能够取得良好的局部编辑效果，但是对于控制点的选择有较高要求，通常需要人工选定。相比之下，我们的方法能够自动化地生成多尺度控制点，支持对于目标物体在多颗粒度下的精准局部编辑。
% Traditional mesh geometry editing typically involves direct manipulation of mesh elements such as vertices, edges, and faces, which is often labor-intensive and time-consuming.
Classical approaches to mesh editing typically rely on the direct manipulation of low-level geometric primitives, such as vertices, edges, and faces. While this allows for fine control over surface geometry, it is often tedious, time-consuming, and requires significant manual effort, especially for complex shapes or large-scale edits.

% To improve usability, free-form deformation methods, such as lattice- or cage-based approaches~\cite{10.1145/800031.808573, jakab2021keypointdeformer}, partition the space around the mesh into controllable regions. These allow users to perform local deformations more efficiently by adjusting a small number of keypoints. However, such methods generally operate at a single spatial scale, limiting their ability to perform fine-grained local edits across multiple levels of detail.
To alleviate this burden, free-form deformation (FFD) techniques introduce an abstraction layer by embedding the mesh within a spatial structure, commonly a lattice or a cage, that defines a deformation domain~\cite{10.1145/800031.808573, jakab2021keypointdeformer}. Users can deform the mesh indirectly by moving a small number of control points associated with this structure. Although such methods improve efficiency and usability, they typically operate at a single spatial scale, limiting their ability to perform fine-grained local edits across multiple levels of detail.

Another class of approaches, handle-based editing methods such as Laplacian Editing~\cite{10.1145/1057432.1057456} and As-Rigid-As-Possible (ARAP) deformation~\cite{ARAP_modeling:2007}, aim to preserve local geometric properties during deformation by optimizing the mesh with respect to user-defined control points. While effective in maintaining shape coherence, these methods rely heavily on careful manual selection of handles, making them less suitable for large-scale or automatic editing tasks.

In contrast, our method automatically generates a hierarchical set of control points, enabling precise and flexible local editing across multiple levels of granularity, without requiring manual specification of deformation handles.

% \begin{equation}
%   \frac{2h}{\uppi}\int_{0}^{\infty}\frac{\sin\left( \omega\delta \right)}{\omega}
%   \cos\left( \omega x \right) \dd\omega = 
%   \begin{cases}
%     h, & \left| x \right| < \delta, \\
%     \frac{h}{2}, & x = \pm \delta, \\
%     0, & \left| x \right| > \delta.
%   \end{cases}
% \end{equation}

\section{Methodology}

%我们提出一种多层级代理的参数化3D表征，利用多层级代理点及其上存储的参数来精确、紧凑地表达目标物体
% 具体而言，给定一个目标3D物体$S$（可以是多模态例如mesh，点云，隐式场。。。），我们通过构建多层级代理点来表达3D物体
% \begin{equation}
%     \mathbf{S}\sim  (\mathbf{C}^{(1)}, \mathbf{C}^{(2)}, \dots, \mathbf{C}^{(L)})
% \end{equation}
%其中，Ci=\{Pi,Fi\}。Pi代表代理点的位置用于表达3D物体的几何，而Fi代表代理点上的特征来表达3D物体的纹理。技术细节会在接下来的sections里展示
% 下标：$R_{1}$
% 下标：$R^{1}$
% {}：$\{\}$
% ~：$\sim$
% 矩阵：粗体大写；向量：粗体小写；标量：小写
% $\mathbf{C}=(\mathbf{C}^{(1)},\mathbf{C}^{(2)},...,\mathbf{C}^{(L)})$

\begin{figure*}[h]
  \centering
  \includegraphics[width=1.0\linewidth]{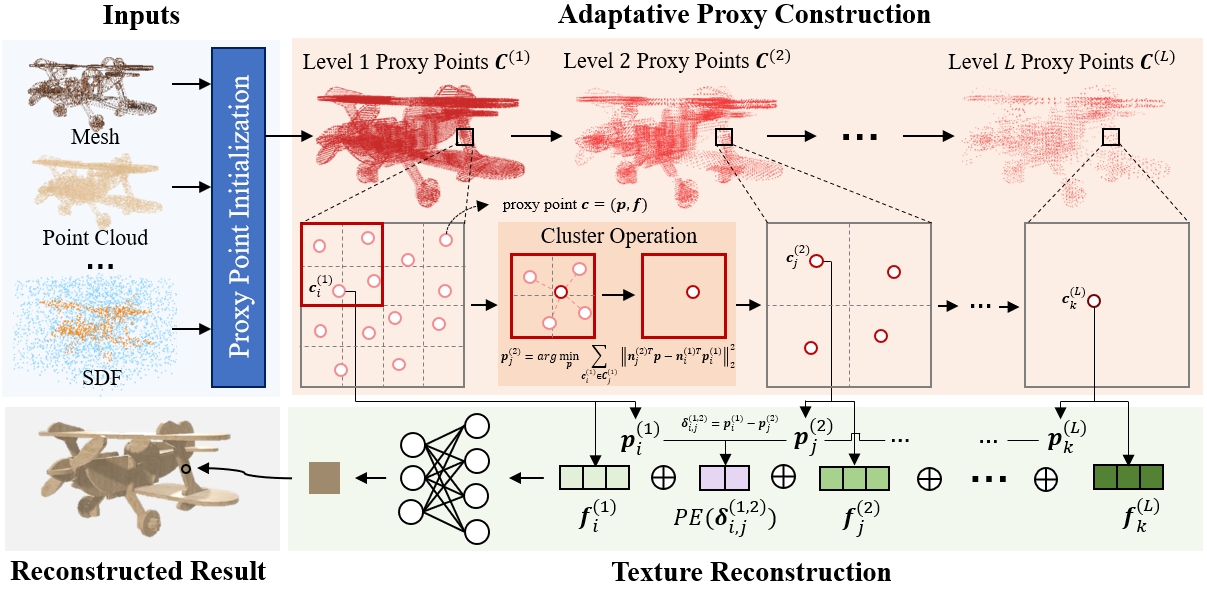}
  \caption{\textbf{Overview of our framework.} We propose a parameterized representation based on a hierarchical structure of proxy points, which enables accurate and efficient reconstruction of 3D objects while supporting multi-scale geometry and texture editing. Given a 3D object in any modality, we first reconstruct its mesh using an existing mesh reconstruction algorithm. The vertices of the reconstructed mesh are then used to initialize the bottom-level proxy points, from which a multi-level proxy hierarchy is constructed via a clustering operation. Texture features are assigned to proxy points at each level. These features, concatenated with positional embeddings, are fed into a decoding function to predict the RGB color of each mesh vertex, enabling high-quality and controllable texture reconstruction and editing.}
  % 我们提出了一种基于多层级代理点构建的参数化表征，能够高精度、高效地重建目标 3D 物体，并支持多粒度的几何编辑和纹理编辑。首先，给定任意模态的 3D 物体，我们先利用 mesh 重建算法重建出 mesh 结构。然后，我们以重建出的 mesh 的顶点为初始化，构建出多层级的代理点集，各层级之间的构造关系利用一个 cluster operation 实现。随后，我们在各层级代理点上存储 texture features。这些 feature 与 positional embedding 拼接以后，会利用一个 decoding function 解码成对应 vertex 的 rgb 颜色，实现高质量且高可控的纹理编辑。
  \label{fig: pipeline}
  % \vspace{-4mm}
\end{figure*}

\subsection{Overview}
% Our framework is illustrated in Fig.~\ref{fig: pipeline},we propose a multi-level proxy-based parametric 3D representation, leveraging multi-level proxy points and the parameters stored on them to precisely and compactly express the target object. Specifically, given a target 3D object $\mathbf{S}$ (which can be multimodal, such as mesh, point cloud, implicit field, \emph{etc.}), we express the 3D object by constructing $L$ levels of proxy points and a xxx function $\theta$ as follows:
% \begin{equation}
%     \mathbf{S}\sim ((\mathbf{C}^{(1)}, \mathbf{C}^{(2)}, \dots, \mathbf{C}^{(L)}), \theta)
% \end{equation}
% where $\mathbf{C}^{(l)}=\{\mathbf{c}^{(l)}_{1}, \mathbf{c}^{(l)}_{2}, ..., \mathbf{c}^{(l)}_{n_l}\}$ denotes the $l$-th level proxy points, $n_l$ is the number of points in $\mathbf{C}^{(l)}$. $\mathbf{c}^{(l)}_i = (\mathbf{p}^{(l)}_i, \mathbf{f}^{(l)}_i)$ represents a $l$-th level proxy point,
% where $\mathbf{p}^{(l)}_i$ is the coordinate of the proxy point, used to represent the geometry of the 3D object. $\mathbf{f}^{(l)}_i$ is the feature on the proxy point, used to represent the texture of the 3D object.
Our proposed framework is depicted in Fig.~\ref{fig: pipeline}. We introduce a multi-level proxy-based parametric 3D representation that utilizes multiple levels of proxy points and the associated features stored on them to represent the target object in a precise and compact manner.
% Specifically, given a target 3D object $\mathbf{S}$ (which can be multimodal, including formats such as mesh, point cloud, implicit field, \emph{etc.}), we express the 3D object by constructing $L$ levels of proxy points along with a decoding function with parameters $\theta$ which decodes textures in a feature-to-rgb manner as follows:
Specifically, given a target 3D object $\mathbf{S}$ (which can be multimodal, such as mesh, point cloud, implicit field, \emph{etc.}), we model the object using $L$ levels of proxy points and a decoding function parameterized by $\theta$ that maps features to RGB textures, as follows:
\begin{equation}
\mathbf{S} \sim ((\mathbf{C}^{(1)}, \mathbf{C}^{(2)}, \dots, \mathbf{C}^{(L)}), \theta).
\end{equation}
Here, $\mathbf{C}^{(l)} = \{\mathbf{c}^{(l)}_1, \mathbf{c}^{(l)}_2, \dots, \mathbf{c}^{(l)}_{n_l}\}$ denotes the set of proxy points at the $l$-th level, where $n_l$ represents the number of points in $\mathbf{C}^{(l)}$. Each proxy point $\mathbf{c}^{(l)}_i = (\mathbf{p}^{(l)}_i, \mathbf{n}^{(l)}_i,\mathbf{f}^{(l)}_i)$ at level $l$ consists of the coordinate $\mathbf{p}^{(l)}_i \in \mathbb{R}^3$ and the normal vector $\mathbf{n}^{(l)}_i \in \mathbb{R}^3$, which are used to represent the geometry of the 3D object, and the feature $\mathbf{f}^{(l)}_i$, which represents the texture of the 3D object. More technical details will be explained in the following sections.

% 这个要改成之前说的那种公式形式
% 这句挺好的，但和前面最好衔接一下
% The key innovation lies in the hierarchical proxy structure, which simplifies traditional vertex-level manipulations into proxy-level control, preserving geometric fidelity while boosting interaction efficiency. 

\subsection{Adaptive Proxy Reconstruction}
% In this section, we propose an adaptive proxy points construction method based on multi-level spatial partitioning and residual-guided clustering. This method takes the triangular mesh generated in surface reconstruction tasks as input and organizes and clusters the original point set at different spatial scales to generate a spatially adaptive proxy point system. It also establishes an extensible hierarchical control structure.
%在本节中，我们提出了一种基于多层次空间划分和误差引导聚类的自适应代理点构建方法。给定多模态输入S，我们的目标是构建L层的代理点C(1)到C(L)，从而生成一个空间自适应的代理点系统，并建立一个可扩展的层次控制结构用来表达S。
%为了捕捉C在不同空间尺度上的几何变化，我们在三维归一化立方体空间中进行L层八叉树网格划分以提取L层代理点，高层级的代理点可以由低层级的代理点推理得出。具体而言，给定第l层代理点集合C（l），我们对空间进行分辨率为2^L-l的八叉树网格划分，定义Grid j为划分后得到的第j个非空网格，它包含部分C(l)内的点集hat{c}(l)_j。
%完成空间划分后，在每个网格单元内，我们利用该区域内点的估计法向量构建基于误差的拟合模型，产生层级为l+1的代理点C（l+1）。具体而言，对于每个 c(l)_k ∈ hat{c}(l)_j，我们假设它大致位于由一个未知点 c ∈ R3 定义的平面上，并为可能的代理点 c(l+1)_k 定义最小二乘优化目标如下：
%minh X c(1)k ∈g_(d) in⊤j h − n⊤j pj^2
%求解该目标可以得到拟合中心 p(l+1)_j 及其对应的误差 r(l+1)_j。由公式（3）可以得知，在包含多点的大网格或曲率变化剧烈的地方，聚类误差大并且聚类程度低，会产生代理点，更多这一机制形成了一个基于曲率的聚类准则，其中误差 r(l+1)_j 有效地刻画了点簇的局部曲率复杂性。如果 r(l+1)_j < ϵ(ϵ是我们设定的聚类误差阈值)   ，我们标记c(l+1)_j为l+1层的有效代理点，它将代理c(l)_k ∈ hat{c}(l)_j，法向量由被代理点取平均后归一化生成。反之，我们直接将 c(l)_k ∈ hat{c}(l)_j作为第l+1层的代理点，它将代理第l层与它们坐标相同的代理点，法向量则与被代理点保持一致。给定多模态输入S，我们首先重建它的几何为mesh，我们称之为M=（V，F）。V是M的顶点集，我们将V作为L1进行初始化，由上面的递推关系得到C（1）-C（L）的多层级代理点。

In this section, we propose an adaptive proxy point construction method based on multi-level spatial partitioning and error-guided clustering. Given a multimodal input $S$, our objective is to construct $L$ layers of proxy points from $\mathbf{C}^{(1)}$ to $\mathbf{C}^{(L)}$, thereby generating a spatially adaptive proxy point system and establishing a scalable hierarchical control structure to represent $S$.

To capture the geometric variations of $C$ at different spatial scales, we perform L-level octree grid partitioning in a three-dimensional normalized cubic space to extract the $L$-level proxy points. Higher-level proxy points can be inferred from lower-level proxy points. Specifically, given the proxy point set $\mathbf{C}^{(l)}$ at the $l$-th level, we partition the space using an octree grid with a resolution of $2^{R-l+1}$, where $R$ is the largest resolution index of the grid. Denote grid $j$ as the $j$-th non-empty grid obtained after partitioning, which contains a subset $\hat{\mathbf{C}}^{(l)}_j$ of points from $\mathbf{C}^{(l)}$
After completing the spatial partitioning, within each grid cell, we construct an error-based fitting model using the estimated normal vectors of the points in that region, generating proxy points $\mathbf{C}^{(l+1)}$
at the $l+1$ level. Specifically, for each $\mathbf{c}^{(l)}_k \in \hat{\mathbf{C}}^{(l)}_j$, we assume $\mathbf{c}^{(l)}_k$ approximately locating on a plane defined by an unknown point $c \in \mathbb{R}^3 $, and define the least squares optimization objective for the possible proxy point $\mathbf{c}^{(l+1)}_j$ as follows:
\begin{equation}
\min_{\mathbf{c}} \sum_{\mathbf{c}^{(l)}_k \in \hat{\mathbf{C}}^{(l)}_j} ( {\mathbf{n}^{(l)}_k}^{\top} \mathbf{c} - {\mathbf{n}^{(l)}_k}^{\top} \mathbf{p}^{(l)}_k )^2.
\label{equ:cluster}
\end{equation}

% Solving this objective yields the fitting center $\mathbf{p}^{(l+1)}_j$ and its corresponding error $\mathbf{r}^{(l+1)}_j$. From Equ.~\ref{equ:cluster}, it can be observed that in large grids containing many points or regions with sharp curvature variations, the clustering error is large and the degree of clustering is low, which leads to the formation of proxy points. This mechanism forms a \textbf{curvature-based clustering criterion} (CBCC), where the error $\mathbf{r}^{(l+1)}_j$ effectively captures the local curvature complexity of the point cluster. For a fitting center $\mathbf{c}^{(l+1)}_j$, if $\mathbf{r}^{(l+1)}_j$ is lower than a threshold $\epsilon$, $\mathbf{c}^{(l+1)}_j$ is marked as an effective proxy point at level $l+1$, which proxies $\mathbf{c}^{(l)}_k \in \hat{\mathbf{C}}^{(l)}_j$, and the normal vector is generated by averaging the normal vectors of the proxied points and normalizing them. In contrast, we directly treat $\mathbf{c}^{(l)}_k \in \hat{\mathbf{C}}^{(l)}_j$ as a proxy point at level $l+1$, which proxies the points at level $l$ with the same coordinates, and the normal vector remains consistent with the proxied points.

Solving this objective yields the fitting center $\mathbf{p}^{(l+1)}_j$ and its associated fitting error $\mathbf{r}^{(l+1)}_j$. As shown in Equation~\ref{equ:cluster}, large clustering errors typically arise in grid regions containing a high density of points or exhibiting sharp curvature variations. In such cases, the degree of clustering is low, and new proxy points are introduced. This behavior gives rise to a \textbf{curvature-based clustering criterion (CBCC)}, where the fitting error $\mathbf{r}^{(l+1)}_j$ serves as an indicator of the local geometric complexity of each point cluster.

For a candidate fitting center $\mathbf{c}^{(l+1)}_j$, if the error $\mathbf{r}^{(l+1)}_j$ is below a threshold $\epsilon$, it is accepted as a valid proxy point at level $l+1$, representing all points $\mathbf{c}^{(l)}_k \in \hat{\mathbf{C}}^{(l)}_j$. The corresponding normal vector is computed by averaging the normals of the proxied points and normalizing the result.
Conversely, if the error exceeds the threshold, each point $\mathbf{c}^{(l)}_k \in \hat{\mathbf{C}}^{(l)}_j$ is retained as an individual proxy point at level $l+1$, inheriting both position and normal direction from the original point.

Given the multimodal input $S$, we first reconstruct its geometry as a mesh, denoted as $\mathcal{M} = (\mathbf{V}, \mathbf{F})$, where 
$\mathbf{V}$ is the set of vertices. We initialize 
$\mathbf{V}$ as ${\mathbf{C}}^{(1)}$, and from the recursive relations above, we obtain the multi-level proxy points from ${\mathbf{C}}^{(1)}$
to ${\mathbf{C}}^{(L)}$.

\subsection{Texture Reconstruction}
%我们为代理点$\mathbf{c}^{(l+1)}_i$设置特征$\mathbf{f}^{(l)}_i$，通过层之间的代理关系并结合位置编码，把高层级的代理点的特征传播到低层级代理点。具体来说，给定$\mathbf{c}^{(l+1)}_k$,它是$\mathbf{c}^{(l)}_j$的代理点，我们可以得到$\mathbf{c}^{(l)}_j$的最终特征\hat{\mathbf{f}}^{(l)}_j：
% \begin{equation}
% \hat{\mathbf{f}}^{(l)}_j = \mathbf{f}^{(l+1)}_k \oplus \mathbf{f}^{(l)}_j \oplus PE\left(\boldsymbol{\delta}^{(l,l+1)}_{j,k}\right)
% \end{equation}
%其中，$\boldsymbol{\delta}^{(l,l+1)}_{j,k} = \boldsymbol{p}^{(l)}_j - \boldsymbol{p}^{(l+1)}_k

%特征按照如上所述的方式从${\mathbf{C}}^{(L)}$逐层传播到${\mathbf{C}}^{(1)}$，$\hat{\mathbf{f}}^{(1)}_i$通过a decoding function $\boldsymbol{\phi}_{\theta}$ parameterized by $\theta$后，将融合特征解码为3D物体的纹理特征（比如rgb,normal,metallicroughness）

%为了获得更精确的纹理重建结果，我们使用多视角渲染损失来优化网络。损失函数包括rgb损失$\mathcal{L}_{\text{rgb}}$和其他纹理属性损失$\mathcal{L}_{\text{others}}$，其形式为：

%在参数优化完成后，我们能够高效地通过编辑不同层级上代理点上的特征，实现对3D物体的纹理编辑，更多细节请参阅接下来的section.

After the adaptive proxy reconstruction, we assign a texture feature $\mathbf{f}^{(l)}_i \in \mathbb{R}^{F^{(l)}}$ to each proxy point $\mathbf{c}^{(l)}_i$ in level $l$. By leveraging the hierarchical proxy correspondence between layers and incorporating positional encoding, we propagate features from higher-level proxy points to those at lower levels. Specifically, given that $\mathbf{c}^{(l+1)}_k$ is a proxy for $\mathbf{c}^{(l)}_j$, the final feature of $\mathbf{c}^{(l)}_j$ can be computed as:
\begin{equation}
\hat{\mathbf{f}}^{(l)}_j = \mathbf{f}^{(l+1)}_k \oplus \mathbf{f}^{(l)}_j \oplus PE\left(\boldsymbol{\delta}^{(l,l+1)}_{j,k}\right),
\end{equation}
where $\boldsymbol{\delta}^{(l,l+1)}_{j,k} = \boldsymbol{p}^{(l)}_j - \boldsymbol{p}^{(l+1)}_k$ denotes the relative positional offset between the proxy point pairs.

The features are propagated in this manner from the topmost layer ${\mathbf{C}}^{(L)}$ down to the bottom layer ${\mathbf{C}}^{(1)}$. The fused feature at the first layer, $\hat{\mathbf{f}}^{(1)}_i$, is then decoded into texture attributes of the 3D object (such as RGB color, surface normals, or metallic-roughness) via a decoding function $\boldsymbol{\phi}_{\theta}$ parameterized by $\theta$.

\subsubsection{Texture Parameters Optimization}

To achieve high-fidelity texture reconstruction, we optimize the network using a multi-view rendering loss. The loss function comprises an RGB reconstruction term $\mathcal{L}_{\text{rgb}}$ and auxiliary texture attribute loss $\mathcal{L}_{\text{others}}$, formulated as,
\begin{equation}
\mathcal{L}_{\text{render}} = \mathcal{L}_{\text{rgb}} + \lambda \mathcal{L}_{\text{others}}.
\end{equation}

After optimization, texture editing of 3D objects can be efficiently performed by modifying features at different proxy levels. Please refer to the following section for more implementation details.
% \begin{equation}
% \mathcal{L}_{\text{render}} = \lambda_{\text{color}} \mathcal{L}_{\text{color}} + \lambda_{\text{normal}} \mathcal{L}_{\text{normal}}
% \end{equation}

% During training, the Adam optimizer is used to adjust the model parameters, improving the quality of texture reconstruction by minimizing the rendering loss.

\subsection{Controllable Editing}
% 由于我们用层级代理点的方式表征目标物体，因此我们可以通过控制代理点的位置和纹理特征来实现目标物体几何和纹理的编辑。
% 正如之前提到，我们将$\mathcal{M} = (\mathbf{V}, \mathbf{F})$的顶点集合$\mathbf{V}$作为底层代理点集合${\mathbf{C}}^{(1)}$。
% 通过构建的层级代理关系，高层级代理点可以代理更多的${\mathbf{c}}^{(1)_i}$。因此无论是几何编辑还是纹理编辑，当我们需要大范围编辑时，我们可以只编辑少数的高层级代理点代理点的位置或特征，而当我们需要进行小范围的精确编辑时，我们再编辑低层级代理点，以此来带动${\mathbf{C}}^{(1)}$的位置与纹理变化。
% As previously mentioned, we take the vertex set $\mathbf{V}$ of the mesh $\mathcal{M} = (\mathbf{V}, \mathbf{F})$ as the set of lowest-level proxy points, denoted as ${\mathbf{C}}^{(1)}$.
Owing to our hierarchical proxy point representation, both geometry and texture of the target object can be edited by manipulating the positions of proxy points and their associated texture features.
Through the constructed hierarchical proxy relationships, higher-level proxy points can represent a larger number of ${\mathbf{c}}^{(1)}_i$. Therefore, for both geometric and texture editing, when large-scale modifications are required, we can simply edit the positions or features of a small number of high-level proxy points. Conversely, for fine-scale, precise editing, we adjust the lower-level proxy points, thereby inducing changes in the positions and textures of ${\mathbf{C}}^{(1)}$ accordingly.
\subsubsection{Geometry Editing.}
% 目标物体的几何编辑可以通过拖动控制点的方式实现。当我们移动代理点$\mathbf{p}}^{(l)}_i$ 发生位移 $\Delta$，它会牵引它控制下的低层级控制点发生相应的移动。这种移动关系会层层传递到最底层的控制点 ${\mathbf{p}}^{(1)}_j$（即重建的 mesh 的顶点），
%当我们移动代理点$\mathbf{p}}^{(l)}_i$，它会牵引${\mathbf{p}}^{(1)}_j$的位置,若\mathbf{p}}^{(l)}_i的位移是$\Delta$
% \begin{equation}
% {\mathbf{p}}^{(1)\prime}_j={\mathbf{p}}^{(1)}_j+ w^{(l)}_{ji}\Delta
% \end{equation}
%为了使得形变更加平滑与可控，我们定义$w^{(l)}_{ji}$为单指数衰减函数，
% $w_i = \exp\left(-\frac{d^{(l)}_{ji}}{\lambda}\right)$ and 
% $d^{(l)}_{ji} = \| \mathbf{p}^{(1)}_j - \mathbf{p}^{(l)}_i \|$，$lamda$是控制形变效果的超参数。
%得到{\mathbf{p}}^{(1)\prime}_j后，我们应用Laplacian Editing (Sorkine et al. 2004)得到最终的几何编辑效果。
%如此当我们需要大范围地进行几何编辑时，只需改变少数高层级代理点的位置，需要细粒度编辑时则可以挪动低层级代理点。并且我们提供多个不同粒度的层级，因此可以在多尺度下进行可控的几何编辑。
% When we move the proxy point $\mathbf{p}^{(l)}_i$, it influences the position of ${\mathbf{p}}^{(1)}_j$. If the displacement of $\mathbf{p}^{(l)}_i$ is $\Delta$, then
Geometric editing of the target object is achieved by manipulating the positions of hierarchical control points. Specifically, when a proxy point $\mathbf{p}^{(l)}_i$ at level $l$ is displaced by $\Delta$, it induces a corresponding movement in its associated lower-level control points. This displacement propagates recursively through the hierarchy and ultimately affects the positions of the bottom-level point ${\mathbf{p}^{(1)}_j}$, which corresponds to a vertex of the reconstructed mesh, to a new position ${\mathbf{p}}^{(1)\prime}_j$, as described in  Equ.~\ref{equ: delta},
\begin{equation}
{\mathbf{p}}^{(1)\prime}_j={\mathbf{p}}^{(1)}_j+ w^{(l)}_{ji}\Delta,
\label{equ: delta}
\end{equation}
where $w^{(l)}{ji}$ is the influence weight that quantifies the extent to which the displacement of $\mathbf{p}^{(l)}_i$ affects the position of the point $\mathbf{p}^{(1)}_j$.
To achieve smoother and more controllable deformations, we define $w^{(l)}_{ji}$ as an exponential function,
\begin{equation}
    w_{ji} = \exp\left(-\frac{d^{(l)}{ji}}{\tau}\right),
\end{equation}
where $d^{(l)}_{ji} = || \mathbf{p}^{(1)}_j - \mathbf{p}^{(l)}_i ||_2$ denotes the l-2 distance between $\mathbf{p}^{(1)}_j$ and $ \mathbf{p}^{(l)}_i$. Scalar $\tau$ is the temperature parameter that controls the vertex deformation effect.

After obtaining ${\mathbf{p}}^{(1)\prime}_j$, we apply Laplacian Editing (Sorkine et al., 2004) to achieve the final geometric editing effect.
In this way, when large-scale geometric edits are required, we only need to modify the positions of a small number of higher-level proxy points. For fine-grained adjustments, we can instead manipulate the lower-level proxy points. Moreover, we provide multiple hierarchical levels at different granularities, enabling controllable geometric editing across multiple scales.

\begin{figure*}[t]
  \centering
   \includegraphics[width=1.0\linewidth]{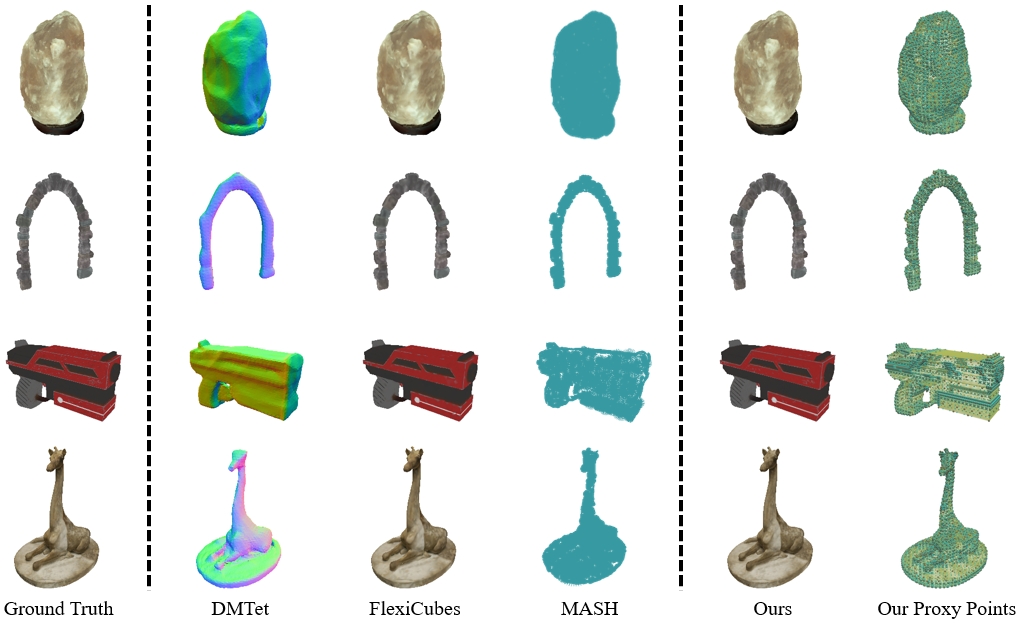}
   % \vspace{-6mm}
   \caption{\textbf{Qualitative comparison on Objaverse dataset.} Qualitative reconstruction results are presented. DMTet~\cite{shen2021deep} and MASH~\cite{li2025mash} show only geometric reconstructions, as they do not support texture modeling. In contrast, our method and FlexiCubes~\cite{shen2023flexible} provide full reconstructions, including both geometry and texture. The last column visualizes the hierarchical proxy point structure produced by our method with $L = 3$ levels, where low- to high-level points are color-coded in yellow, cyan, and brown with increasing point sizes, respectively.}
   % Qualitative reconstruction results are displayed. DMTet~\cite{shen2021deep} 和 MASH~\cite{li2025mash} 只展示了几何重建的结果，因为它们不支持纹理重建。我们的方法和 FlexiCubes~\cite{shen2023flexible} 展示了完整的重建结果（纹理+几何）。最后一列展示了我们方法构建的 $L=3$ 层代理点结构，从低到高分别用褐色、青色、黄色来展示。
   % \vspace{-4mm}
   \label{fig: objaverse}
\end{figure*}

\subsubsection{Texture Edting.}
%我们同样可以通过改变不同层级的代理点特征${\mathbf{f}}^{(l)}_i$实现不同范围的纹理编辑，当需要大范围编辑纹理时可以改变高层级代理点的特征来影响更大区域，当需要更精确的渲染时则需要控制低层级代理点上特征的改变。特别的，我们可以实现指定的同层级代理点之间的特征转移，从而可控地实现3D物体不同区域的纹理迁移。具体而言，我们在空间上刚性对齐两片区域的同层级代理点，通过线性插值的方式将一片区域的代理点特征转移到目标区域代理点，经过decoding function $\boldsymbol{\phi}_{\theta}$后得到迁移后的纹理。
Similarly, we can achieve texture editing at different scales by modifying the features of proxy points at various levels, denoted as ${\mathbf{f}}^{(l)}_i$. For large-scale texture editing, altering the features of higher-level proxy points allows us to affect broader regions. In contrast, fine-grained rendering requires precise control over the features of lower-level proxy points. Notably, we enable controlled texture transfer between different regions of a 3D object by transferring features across proxy points within the same level. Specifically, we rigidly align two regions of same-level proxy points in 3D space and perform linear interpolation to transfer the features from one region's proxy points to those of the target region. The transferred texture is then obtained through the decoding function $\boldsymbol{\phi}_{\theta}$.

\section{Experiments}

\subsection{Experimental Setups}

\subsubsection{Dataset.}
% We conduct experiments on two widely used 3D object datasets: ShapeNet~\cite{chang2015shapenet} and Objaverse~\cite{objaverse}. For ShapeNet, all models are uniformly normalized to fit within a cube of $[-0.9, 0.9]$, and comparisons across methods are performed on these normalized meshes to ensure fairness.
% For Objaverse, we curate a test set of 1,000 structurally well-defined, color-textured objects. All models are preprocessed using the same normalization strategy as in the ShapeNet dataset.
We conduct experiments on Objaverse~\cite{objaverse}, a widely used dataset for 3D generation and reconstruction. We curate a test set consisting of 1,000 structurally well-defined, color-textured objects. To ensure fair comparison across methods, all models are uniformly normalized to fit within a cube of $[-0.9, 0.9]$ before evaluation.

\subsubsection{Implementation Details.}
%在Adaptive Proxy Reconstruction中，我们定义大小为5.0的聚类误差阈值，提取3层代理点。我们将%${\mathbf{f}}^{(1)}_i$，${\mathbf{f}}^{(2)}_i$，${\mathbf{f}}^{(3)}_i$设置为（32，）（24，）（12，）的tensor,设置2层128维的mlp作为decoding function $\boldsymbol{\phi}_{\theta}$。
% We define a clustering error threshold of 5.0 and extract three levels of proxy points and set the feature tensors ${\mathbf{f}}^{(1)}_i$, ${\mathbf{f}}^{(2)}_i$. Then we ${\mathbf{f}}^{(3)}_i$ to have dimensions (32,), (24,), and (12,), respectively. A two-layer MLP with 128-dimensional hidden units is employed as the decoding function $\boldsymbol{\phi}{\theta}$.

% 在实验中，我们构建 $L=3$ 层的控制点，最底层控制点直接用重建出的 mesh 顶点初始化。在 adaptive proxy reconstruction 的过程中，八叉树 grid 的最大分辨率指数 $R=7$，这意味着 grid 的最大分辨率为 $2^R=128$。各层之间的聚类误差阈值设置为 $\epsilon=5.0$。1~3层代理点上存的 texture feature 的维度 $F^{(l)}, l=1, 2, 3$ 分别设置为 $32$, $24$ 和 $12$。Positional embedding 的 dimension 为 $60$。The decoding function $\boldsymbol{\phi}{\theta}$ 通过一个具有两个隐藏层，128通道的全连接MLP网络实现。训练中 $\mathcal{L}_{others}$ 的系数为 $\lambda=0.5$。在几何编辑中，计算 $w_{ji}$ 时用的 temperature parameter $\tau=1.0$。

In our experiments, we construct a control point hierarchy with $L = 3$ levels. The bottom-level control points are initialized directly from the vertices of the reconstructed mesh. During adaptive proxy reconstruction, the maximum resolution exponent of the octree grid is set to $R = 7$, corresponding to a maximum spatial resolution of $2^R = 128$. The clustering error threshold between levels is set to $\epsilon = 5.0$.
The dimensions of the texture features $F^{(l)}$ stored at levels $l = 1, 2, 3$ are set to 32, 24, and 12, respectively. The positional embedding has a dimension of 60. The decoding function $\boldsymbol{\phi}_{\theta}$ is implemented as a fully connected MLP with two hidden layers of 128 channels each.
During the texture feature training, the coefficient for the auxiliary loss term $\mathcal{L}_\text{others}$ is set to $\lambda = 0.5$. For geometry editing, the temperature parameter used in computing the influence weights $w$ is set to $\tau = 1.0$.

\begin{table*}[t]
\small
\centering
\begin{tabular}{l|ccccccc}
\toprule
Method & CD $\downarrow$ & PSNR $\uparrow$ & SSIM $\uparrow$ & \#Params (G) $\downarrow$ & \#Params (G+T) $\downarrow$ & Time (G) $\downarrow$ & Time (G+T) $\downarrow$ \\ 
\midrule
DMTet~\cite{shen2021deep} & $0.0269$ & $-$ & $-$ & $1.1\times10^6$ & $-$ & $\underline{2\text{min}}$ & $-$ \\
FlexiCubes~\cite{shen2023flexible} & $0.0216$ & $\underline{35.55}$ & $\underline{0.9835}$ & $6.6\times10^6$ & $\underline{7.4\times10^6}$ & $5$min & $\boldsymbol{8}$\textbf{min} \\
MASH~\cite{li2025mash} & $\boldsymbol{0.0145}$ & $-$ & $-$ & $\boldsymbol{1.0\times10^5}$ & $-$ & $1.5$h & $-$ \\
Ours & $\underline{0.0207}$ & $\boldsymbol{37.12}$ & $\boldsymbol{0.9858}$ &$\underline{9.7\times10^5}$  & $\boldsymbol{2.6\times10^6}$ & $\boldsymbol{30}$\textbf{s} & $\underline{27\text{min}}$ \\ 
\bottomrule
\end{tabular}
% \caption{\textbf{Geometry and appearance reconstruction results on Objaverse Dataset.} The L2 chamfer distances (CD) for geometry evaluation, PSNR for appearance evaluation, and parameter usage (\#Params) of both methods are reported. The PSNR is calculated over $50$ random views which are consistent across all methods. Our method outperforms baselines on both geometry and appearance reconstruction tasks.}
% \vspace{-2mm}
% \caption{\textbf{Geometry and appearance reconstruction results on Objaverse.}
% We report L2 Chamfer Distance (CD) for geometry evaluation, and PSNR (dB) and SSIM for appearance evaluation. \#Params ($\times10^5$) lists the number of parameters for geometry only and joint geometry + texture representation (separated by a slash). For methods without texture modeling (DMTet and MASH), only the geometry parameter count is provided. PSNR and SSIM are computed over 50 consistent random views across all methods.}
\caption{\textbf{Geometry and appearance reconstruction results on Objaverse.}
We report L2 Chamfer Distance (CD) for evaluating geometric accuracy, and PSNR (dB) and SSIM for assessing texture reconstruction quality. The \#Params columns indicate the number of parameters used for geometry only (G) and for joint geometry and texture representation (G+T). The Time columns report the inference time for geometry-only reconstruction (G) and joint reconstruction (G+T). For methods that do not support texture modeling (e.g., DMTet and MASH), only the geometry parameter count and optimization time are reported.
% PSNR and SSIM are computed over 50 randomly sampled, consistent views across all methods.
}
\label{tab: objaverse}
% \vspace{-6mm}
\end{table*}

\subsection{Comparison on 3D Reconstruction}
% 我们在 3D 重建任务上对比我们的方法与 SOTA 的 3D 表征，包括 DMTet~\cite{shen2021deep}, FlexiCubes~\cite{shen2023flexible} 和 Mash~\cite{li2025mash}。具体而言，我们比较我们的方法与对比方法在几何和纹理（如果支持纹理重建）重建方面的定量与定性表现。几何重建的定性表现用 Chamfer distance 来衡量；纹理重建的定性表现用 PSNR 和 SSIM 来衡量。All baselines use official implementations with default hyperparameters.
We evaluate our method on the 3D reconstruction task and compare it against several state-of-the-art 3D representations, including DMTet~\cite{shen2021deep}, FlexiCubes~\cite{shen2023flexible}, and MASH~\cite{li2025mash}, a concurrent 3D representation work. Specifically, we assess both the geometric and texture reconstruction performance (where applicable), using a combination of quantitative and qualitative metrics.
For geometry, reconstruction quality is measured using Chamfer Distance (CD). For texture, we report PSNR and SSIM~\cite{wang2004image} to evaluate fidelity. All baselines are evaluated using their official implementations with default hyperparameter settings, unless otherwise specified.

Notably, we observed that when using the default configuration provided in the official implementation, MASH tends to suffer from frequent early stopping on the Objaverse dataset, resulting in severe underfitting. To ensure a fair comparison, we increased the number of anchor points in MASH to $4000$ and set the early stopping parameter ``min\_delta" to $1\times 10^{-8}$. This adjustment effectively mitigates the underfitting issue and allows MASH to achieve strong reconstruction performance on Objaverse. However, it also significantly increases the computational cost and reconstruction time of the representation.

% Quantitative results are listed in Tab.~\ref{tab: objaverse}. PSNR 和 SSIM 指标是在 $50$ 个随机采样视角下测量出的。优化时间指标 (Time) 是在单张 RTX 3090 显卡上测试得到的。As we can see, 相比于 DMTet 和 FlexiCubes，我们在几何重建质量 (chamfer distance) 和纹理重建质量 (PSNR 和 SSIM) 上都取得了显著的提升。不仅如此，我们的方法在重建几何（以及纹理）时都比这两个对比方法使用更少的参数，体现出我们方法在重建目标物体时的紧凑性。MASH 表征能够用很少的参数量表征出更好的几何效果，但是正如我们之前所说，MASH 的优化对于超参数十分敏感，错误的超参数设置会导致其优化不稳定。除此以外，MASH 在 Objaverse 数据集上只重建几何需要约 $1.5$h 的优化时间，相比而言我们的方法只需要 $30$s。再者，MASH 方法在几何上的高度紧凑性使其难以表征纹理，而我们的多层级控制点能很好地解决这个问题。
Quantitative results are presented in Tab.~\ref{tab: objaverse}. PSNR and SSIM are computed over 50 randomly sampled viewpoints, while the optimization time (Time) is measured on a single NVIDIA RTX 3090 GPU.
As shown, our method achieves significant improvements in both geometric reconstruction quality (CD) and texture reconstruction quality (PSNR and SSIM) compared to DMTet and FlexiCubes. Furthermore, our approach requires fewer parameters for both geometry and joint geometry-texture reconstruction, demonstrating its compactness and efficiency in representing 3D objects.

While MASH achieves strong geometric accuracy with a small number of parameters, it is highly sensitive to hyperparameter settings, and suboptimal configurations often lead to unstable optimization. In addition, MASH requires approximately 1.5 hours to optimize geometry on the Objaverse dataset, whereas our method completes the joint geometry and texture reconstruction in just 30 seconds.
Moreover, although MASH achieves compact geometric representation, its design makes it unsuitable for texture reconstruction. In contrast, our hierarchical control point representation effectively supports both accurate geometry and high-fidelity texture modeling.

% Qualitative reconstruction results are shown in Fig.~\ref{fig: objaverse}. 可以看到，相比于另一个支持纹理重建的表征 FlexiCubes，我们的方法能够重建出更清晰的纹理。这得以与我们的方法使用了多层级的纹理特征+基于MLP的decode方式，这使得我们能够表征更加细粒度的纹理细节。
Qualitative reconstruction results are shown in Fig.~\ref{fig: objaverse}. Compared to FlexiCubes, the only other method that supports texture reconstruction, our approach produces noticeably sharper and more detailed textures. This improvement is attributed to our use of multi-level texture features combined with an MLP-based decoding scheme, which enables the representation of fine-grained appearance details.

\subsection{Easy 3D Editing}
% As mentioned in \textbf{Methodology}, 我们的方法支持通过操控代理点的方式实现 3D 物体的可控几何和纹理编辑，下面我们分别展示我们的表征在这两个功能下的定性表现。
As discussed in \textbf{Methodology}, our representation supports controllable 3D geometry and texture editing via manipulation of proxy points. In the following, we present qualitative results to demonstrate the effectiveness of our method in both editing scenarios.

\subsubsection{Shape Editing.}
% 我们的方法支持通过简单拖动代理点的方式对原始物体的形状进行编辑。不仅如此，由于我们的方法构建了多层级代理点，不同层级的代理点能够影响不同范围大小的物体表面区域（越高层的代理点影响的区域越大），因此通过拖拉不同层级的代理点，我们的表征支持不同粒度下的物体编辑效果，这是传统的 mesh 编辑方法无法实现的。
Our method enables intuitive shape editing of 3D objects by simply dragging proxy points.
More importantly, due to the hierarchical structure of our proxy point representation, proxy points at different levels influence surface regions of varying spatial extents—higher-level points affect broader areas, while lower-level points offer finer control.
This design allows for multi-scale geometry editing by manipulating proxy points at different levels, which is difficult to achieve using traditional mesh editing techniques.

\begin{figure}[t]
  \centering
   \includegraphics[width=1.0\linewidth]{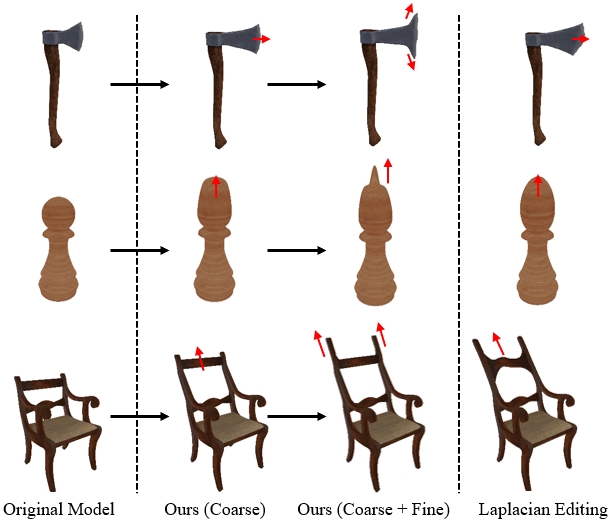}
   % \vspace{-6mm}
   \caption{\textbf{Geometry editing results.} We compare the geometric editing capabilities of our method with Laplacian editing~\cite{10.1145/1057432.1057456}. The second and third columns illustrate coarse- and fine-grained edits achieved by manipulating high-level and low-level proxy points, respectively. Such multi-scale, part-aware editing is not feasible using traditional mesh editing techniques like Laplacian editing.}
   % \textbf{Geometry editing results.} 这里我们对比了我们的方法和 Laplacian Editing~\cite{10.1145/1057432.1057456} 在几何编辑上的表现。第二列和第三列分别展示通过拖动高层级代理点和低层级代理点实现的不同粒度的几何编辑。这是传统 mesh 编辑方法（例如 Laplacian Editing，第四列）所无法做到的。
   \label{fig: geometry editing}
   % \vspace{-4mm}
\end{figure}

% 我们对比了我们的编辑方法与传统 mesh 编辑方法 Laplacian editing~\cite{10.1145/1057432.1057456} 在几何编辑上的定性表现。Laplacian editing 是通过 Blender 中的相应工具实现的，实现过程设计大量的人工操作。实验结果如 Fig.~\ref{fig: geometry editing} 所示。图中的第二列的结果是我们的方法通过拖动高层 (Level 3) 代理点得到的大尺度编辑结果（例如把斧子变大或把椅背拉长）；第三列则是在第二列结果的基础上，再拖动底层控 (Level 2) 代理点得到的小尺度编辑结果（例如进一步微调椅子背上的支撑框架）。可以看到我们的方法能够在仅使用数次拖动的情况下实现各种粒度的形状编辑。相比之下，Laplacian editing 的编辑过程不仅涉及大量的点击操作（例如区域选的、控制点挑选等）作为前序步骤，并且每次仅能实现单一尺度下的几何形变。多尺度操作需要重新进行繁琐的前序操作。
We compare the qualitative geometry editing performance of our method with the traditional mesh editing technique, Laplacian editing~\cite{10.1145/1057432.1057456}. The Laplacian editing baseline is implemented using the corresponding tool in Blender, which involves extensive manual operations during the editing process.
The comparison results are shown in Fig.~\ref{fig: geometry editing}. The second column displays large-scale edits produced by our method through manipulation of high-level (Level 3) proxy points, such as enlarging the blade of an axe or extending the backrest of a chair. The third column shows fine-scale adjustments obtained by further manipulating low-level (Level 2) proxy points based on the previous edits, such as refining the supporting frame on the chair back. As shown, our method supports flexible and multi-scale shape editing with only a few intuitive drag operations. In contrast, Laplacian Editing requires extensive manual steps for each deformation, operates at a single scale per edit, and demands repeated re-selection of regions and handles when switching scales.

\subsubsection{Texture Editing.}
%针对3D物体特定区域的精准纹理编辑是一个非常具有挑战性的任务。现有的方法通常面临着纹理扭曲、区域选择困难以及低效率的问题，尤其在处理复杂形状和多尺度纹理时。我们提出的方法通过精确调整少量代理点的特征，使得在复杂的3D物体上实现对特定感兴趣区域的高精度纹理编辑。我们的方法的核心优势在于能够通过操作代理点的局部特征，精确控制纹理的变换与迁移。
%其中一个重要且有趣的做法是通过迁移代理点的特征来传递纹理。这一过程涉及在3D模型中根据代理点的位置信息对纹理进行重新映射，从而实现目标区域的纹理传递，结果如图5所示。具体而言，图中的相邻两列展示了在基于我们代理点表示的3D物体纹理与通过迁移代理点特征后的纹理效果对比。通过这种方式，我们能够在不失真或扭曲的情况下，将任意形状和大小的选定区域的纹理精准地传递到目标区域。这一现象归因于我们方法中的两个关键因素：首先是精确的代理点对齐，其次是基于距离的插值方法，这使得纹理传递过程能够平滑且连续。
Precise texture editing of specific regions on 3D objects is a highly challenging task. Existing methods often encounter issues such as texture distortion, difficulty in region selection, and inefficiency, particularly when handling complex shapes and multi-scale textures. The method we propose enables high-precision texture editing of specific regions of interest on complex 3D objects by precisely adjusting the features of a small number of proxy points. The core advantage of our method lies in its ability to manipulate the local features of proxy points, allowing for accurate control over texture transformation and transfer.

\begin{figure}[t]
  \centering
   \includegraphics[width=1.0\linewidth]{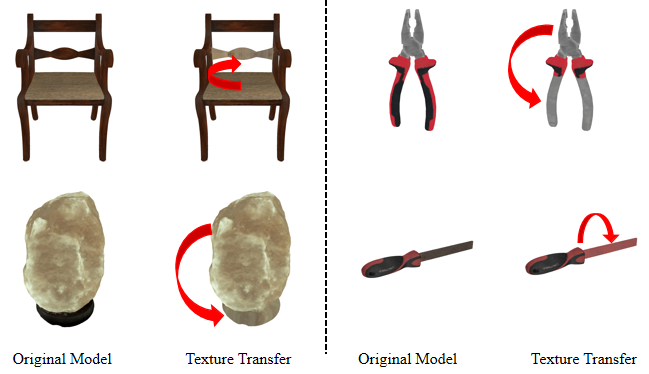}
   % \vspace{-6mm}
   \caption{\textbf{Texture editing results.} Comparison of texture transfer results. The adjacent columns show the texture of a 3D object based on our proxy point representation (left) and the texture effect after migrating the proxy point features (right). Regardless of shape or size, the precise transfer of texture to the target region is achieved through accurate proxy point alignment and distance-based interpolation.}
   % \textbf{Geometry editing results.} 这里我们对比了我们的方法和 Laplacian Editing~\cite{10.1145/1057432.1057456} 在几何编辑上的表现。第二列和第三列分别展示通过拖动高层级代理点和低层级代理点实现的不同粒度的几何编辑。这是传统 mesh 编辑方法（例如 Laplacian Editing，第四列）所无法做到的。
   \label{fig: texture editing}
   % \vspace{-4mm}
\end{figure}

One important and interesting approach is the transfer of texture by migrating the features of proxy points. This process involves remapping the texture based on the positions of the proxy points in the 3D model, enabling the transfer of texture to the target region, as shown in Fig.~\ref{fig: texture editing}. The adjacent columns in the figure display the texture of the 3D object based on our proxy point representation and the texture effect after migrating the proxy point features. Through this approach, we are able to precisely transfer the texture of selected regions, regardless of their shape or size, to the target area without distortion or warping. This phenomenon can be attributed to two key factors in our method: first, the precise alignment of proxy points, and second, the distance-based interpolation technique, which ensures that the texture transfer process remains smooth and continuous.

\subsection{Ablation Study}
\subsubsection{Component Analysis.}
We perform an ablation study to assess the impact of key components in our framework, including the positional embedding, curvature-based clustering criterion, and multi-level texture features. For simplicity and efficiency, experiments are conducted on a selected subset of the Objaverse dataset. We use PSNR and SSIM as quantitative metrics to evaluate texture reconstruction performance. The quantitative results are reported in Table~\ref{tab: ablation}.

% As shown, 去掉任意一个 component 都会导致渲染效果的下降。在三个 component 中，positional embedding 是对结果贡献最大的组件，因为它为 decoding function $\phi(\theta)$ 带来了准确地相邻层级代理点之间的相对位置信息，方便 $\phi(\theta)$ 更准确地融合各层级的 texture feature；Curvature-based clustering criterion 的使用能够使表征在模型表面变换剧烈的地方自适应地放置更多的代理点，进而更好地表征这些复杂的区域；Multi-level texture features 的策略最初是用于实现多层级纹理编辑的，但我们发现它对于提高纹理重建质量也有显著的帮助。
As shown, removing any individual component leads to a noticeable degradation in rendering quality. Among the three components, \textbf{positional embedding} contributes the most, as it provides the decoding function $\phi_\theta$ with precise relative positional information between proxy points across different levels, enabling more accurate fusion of multi-level texture features.
The \textbf{curvature-based clustering criterion} helps allocate more proxy points in regions with high surface variation, allowing the representation to better capture local geometric complexity. Although the \textbf{multi-level texture feature} design was originally introduced to support hierarchical texture editing, we observe that it also significantly improves the quality of texture reconstruction.

\begin{table}[t]
% \small
\centering
\begin{tabular}{>{\centering\arraybackslash}p{1cm}>{\centering\arraybackslash}p{1cm}>{\centering\arraybackslash}p{1cm}|>{\centering\arraybackslash}p{1.5cm}>{\centering\arraybackslash}p{1.5cm}}
\toprule
PE & CBCC & MLF & PSNR $\uparrow$ & SSIM $\uparrow$ \\
\midrule
$\times$ & \checkmark & \checkmark & $35.2781$ & $0.9751$ \\
\checkmark & $\times$ & \checkmark & $35.3606$ & $0.9756$ \\
\checkmark & \checkmark & $\times$ & $35.6216$ & $0.9757$ \\
\checkmark & \checkmark & \checkmark & $\textbf{35.6897}$ & $\textbf{0.9772}$ \\
\bottomrule
\end{tabular}
% \vspace{-1mm}
\caption{\textbf{Component analyses.} We evaluate the impact of removing individual design components, including the positional embedding (PE), curvature-based clustering criterion (CBCC), and multi-level texture features (MLF), on 3D reconstruction performance using an Objaverse subset. PSNR and SSIM are reported under various ablation settings.
% The full model achieves the best performance, and each component contributes positively to the overall texture reconstruction quality.
}
\label{tab: ablation}
% \vspace{-3mm}
\end{table}

\section{Conclusion}
% This paper presents a hierarchical proxy point-based method for 3D object representation that enables compact and high-precision modeling, along with controllable geometry and texture editing at multiple levels of granularity. Our approach allows for accurate reconstruction and efficient editing, both at large and fine scales, by manipulating proxy points across different levels.The hierarchical structure ensures efficient global changes by adjusting higher-level proxy points for large-scale edits, while fine-grained modifications can be made by adjusting lower-level points. This flexibility significantly enhances the editing process, making it applicable in a wide range of scenarios.Additionally, the adaptive algorithm dynamically adjusts proxy point distribution based on geometric and texture features, optimizing the model construction and reducing computational costs. In summary, our method offers a new, efficient solution for 3D modeling and editing, improving both precision and flexibility. As related technologies evolve, this hierarchical approach has the potential for broader applications.
This paper presents a hierarchical proxy point-based method for 3D object representation that enables compact, high-precision modeling and controllable geometry and texture editing at multiple levels of granularity. By manipulating proxy points across different levels, our approach supports both coarse and fine-grained reconstruction and editing. Higher-level points enable efficient global changes, while lower-level points allow localized refinements, making the editing process more flexible and widely applicable. An adaptive algorithm further optimizes proxy point distribution based on geometric and texture features, improving modeling efficiency and reducing computational cost. Overall, our method provides a novel, precise, and flexible solution for 3D modeling and editing, with strong potential for broader applications as related technologies advance.

\bibliography{aaai2026}

\end{document}